\documentclass{emulateapj}
\usepackage{latexsym,amssymb}
\usepackage{natbib}
\usepackage{hyperref}

\newcommand{\boldsymbol}[1]{\mbox{\boldmath{$#1$}}}

\newcommand{\ddr}[1]{\ensuremath{\frac {\partial {#1}}{\partial r}}}
\newcommand{\ddt}[1]{\ensuremath{\frac {\partial {#1}}{\partial t}}}
\newcommand{\DDt}[1]{\ensuremath{\frac {d {#1}}{d t}}}

\newcommand{\ddphi}[1]{\ensuremath{\frac {\partial {#1}}{\partial \phi}}}

\newcommand{\ddx}[1]{\ensuremath{\frac {\partial {#1}}{\partial x}}}
\newcommand{\ddy}[1]{\ensuremath{\frac {\partial {#1}}{\partial y}}}
\newcommand{\ddb}[1]{\ensuremath{\frac {\partial {#1}}{\partial b}}}

\newcommand{\sigmabar}{\ensuremath{\bar{\sigma}}}
\newcommand{\vorticity}{\ensuremath{\bar{\omega}}}

\newcommand{\uvec}{\ensuremath{\boldsymbol{u}}}
\newcommand{\nablavec}{\ensuremath{\boldsymbol{\nabla}}}
\newcommand{\Omegavec}{\ensuremath{\boldsymbol{\Omega}}}
\newcommand{\xvec}{\ensuremath{\boldsymbol{x}}}
\newcommand{\kvec}{\ensuremath{\boldsymbol{k}}}

\newcommand{\utildevec}{\ensuremath{\tilde{\boldsymbol{u}}}}
\newcommand{\Ptilde}{\ensuremath{\tilde{P}}}
\newcommand{\rhotilde}{\ensuremath{\tilde{\rho}}}

\newcommand{\utildeveceps}{\ensuremath{\tilde{\boldsymbol{u}}_{\epsilon}}}
\newcommand{\Ptildeeps}{\ensuremath{\tilde{P}_{\epsilon}}}
\newcommand{\rhotildeeps}{\ensuremath{\tilde{\rho}_{\epsilon}}}

\newcommand{\Lmatrix}{\ensuremath{\mathfrak{L}}}

\shorttitle{Heavy Vortices}
\shortauthors{Chang \& Oishi}

\begin{document}

\title{On the Stability of Dust-Laden Protoplanetary Vortices} 

\author{Philip Chang}
\affil{Canadian Institute for Theoretical Astrophysics, 60 St George St, Toronto, ON M5S 3H8, Canada} 
\email{pchang@cita.utoronto.ca}
\author{Jeffrey S. Oishi}
\affil{Department of Astronomy, 601 Campbell Hall, University of California, Berkeley, CA 94720-3411}
\email{jsoishi@astro.berkeley.edu}



\begin{abstract}
The formation of planetesimals via gravitational instability of the
dust layer in a protoplanetary disks demands that there be local
patches where dust is concentrated by a factor of $\sim$ a few $\times
10^3$ over the background value. Vortices in protoplanetary disks may
concentrate dust to these values allowing them to be the nurseries of
planetesimals. The concentration of dust in the cores of vortices
increases the dust-gas ratio of the core compared to the background
disk, creating a "heavy vortex."  In this work, we show that these
vortices are subject to an instability which we have called the
heavy-core instability.  Using Floquet theory, we show that this
instability occurs in elliptical protoplanetary vortices when the
gas-dust density of the core of the vortex is heavier than the ambient
gas-dust density by a few tens of percent. The heavy-core instability
grows very rapidly, with a growth timescale of a few vortex rotation
periods.  While the nonlinear evolution of this instability remains
unknown, it will likely increase the velocity dispersion of the dust
layer in the vortex because instability sets in well before sufficient
dust can gather to form a protoplanetary seed.  This instability may
thus preclude vortices from being sites of planetesimal formation.
\end{abstract}

\begin{keywords}
{
accretion, accretion disks -- hydrodynamics -- instabilities
planetary systems: formation -- planetary systems: protoplanetary disks
}
\end{keywords}

\section{Introduction}\label{sec:intro}

Current theories of planet formation postulate that dust grows from
interstellar grain sizes ($\sim \mu m$) to $\sim \mathrm{km}$ size
planetesimals in the disks observed around young stars. This process
must proceed in stages: below $\sim$ cm scales, growth proceeds by
"sticky" grain-grain collisions.  Above $\sim$ km scales, planetesimal
growth proceeds by gravitational accretion.  However, around meter
scales, collisions between grains are destructive and another growth
process is needed. 

This process must be quick. The gaseous disk has a radial pressure
gradient which partially supports it against gravity, leaving its
rotation rate sub-Keplerian. The dust, meanwhile, sees no pressure
gradient and therefore orbits at the Keplerian rate, leading to a
headwind on the dust as it orbits in the gaseous disk. As a result,
the dust rapidly spirals in to the star, at a rate
\begin{equation}
  t \sim 200 \left(\frac{r}{AU}\right)^{13/14} \mathrm{yr}
\end{equation}
in the minimum mass solar nebula--far too short for the formation of
planets \citep[for a recent review, see][]{ChiangYoudin2009}. 

Gravitational instability in the dust layer is one mode by which
planetesimals can grow at the m scale
\citep{Goldreich1973,Safronov1969}.  However, for this to proceed the
dust density must be significantly enhanced by a factor of $\sim$ a
few $\times 10^3$.  The settling of dust to the midplane can enhance
the dust density enormously, but when the dust density, $\rho_{\rm d}$
is similar to the gas density, $\rho_{\rm g}$, Kevin-Helmholtz
instabilities limit further concentration
\citep{Weidenschilling1993,Chiang2008,Barranco2009}.  Hence, the dust
density enhancement is limited to $\sim 100$ for solar metallicity
unless the metallicity of the gas is enhanced (see
\citealt{Youdin2002}) or the dust surface density is enhanced.

A natural way around these somewhat daunting timescale and surface
density problems is to postulate the existence of regions in the disk
where dust is significantly concentrated. This concentration may take
place in persistent, large-scale vortices. On large enough scales
(roughly $\gtrsim H$, where $H = c_s/\Omega$ is the scale height of
the disk, $\Omega = \sqrt{GM_*/r^3}$ is the orbital frequency, $M_*$
is the mass of the star, $r$ is the radius, and $G$ is Newton's
constant), the dynamics of a thin disk ($H/r << 1$) become quasi-two
dimensional. Such flows host inverse cascade processes in which energy
flows to large scales, making the appearance of coherent, long-lived
vortices a distinct possibility. Such large scale vortices could be
seeded by the baroclinic instabilities
\citep{Lovelace1999,Varniere2006,Lyra2009,Lesur2009b}, or by the
quasi-2D decay of initial turbulence generated from the initial
accretion flow from the pre-stellar envelope on to the disk
\citep{Bracco1999}.

Previous studies, both analytic \citep{Barge1995, Tanga1996,
  Chavanis2000} and computational \citep{Bracco1999,Godon2000,
  Lyra2009}, have demonstrated that anticyclonic vortices (those with
vorticity antiparallel to the Keplerian rotation) effectively trap
dust.  As these vortices collect dust, the dust-to-gas ratio in the
vortices increases.  Hence, the gas in the vortices is denser than the
gas in the surrounding disk.  Can a density gradient between the
vortex and surrounding gas or a density gradient within the vortex
trigger new instabilities, destroying or modifying the vortex in the
process? As we will show in this paper, the answer is yes: density
gradients in vortices are destabilizing for vortices sufficiently
heavy cores (and also for all vortices with light cores).

We note that this is not the only means by which dust can be
concentrated into small regions of a protoplanetary disk.
\citet{Youdin2005} have shown that the (inward) radial migration of
dust in a gaseous disk is subject to a gas-dust streaming instability.
The nonlinear evolution of this streaming instability leads to large
concentrations of dust \citep{Youdin2007,Johansen2007}, which might
also be the sites of protoplanetary seed formation.  As our primary
interest is the stability of vortices, we do not study this mechanism,
but mention it for completeness.

This paper is organized as follows.  We begin by discussing
equilibrium solutions for vortices in protoplanetary disks in
\S\ref{sec:vortices}, focusing on the \citet{Kida1981}
(\S\ref{sec:kida}) and \citet[hereafter GNG]{Goodman1987}
(\S\ref{sec:gng}) solutions.  We calculate vortical stability in
\S\ref{sec:stability}, beginning with a description of Floquet theory
(\S\ref{sec:floquet}) and then applying it to our equilibrium vortices in
\S\ref{sec:instability}.  We find two regions where vortices are
unstable: vortices with light cores and vortices with sufficiently
heavy cores.  The growth rate of the instability is quite rapid for a
sufficient density contrast.  We discuss its application to
protoplanetary vortices in \S\ref{sec:discussion} and close with a
summary of our results and discussion of outstanding issues in
\S\ref{sec:conclusions}.


\section{Vortices in Protoplanetary Disks}\label{sec:vortices}

We model a local patch of the disk with a guiding center
radius, $r_0$, with angular velocity $\Omega$ using the incompressible
shearing sheet approximation \citep{Goldreich1965}
\begin{eqnarray}
\ddt{\rho} + u\ddx{\rho} + v \ddy{\rho} &=& 0,\label{eq:continuity}\\
\ddt{u} + u\ddx{u} + v\ddy{u} - 2\Omega v - 3\Omega^2 x&=& -\frac{1}{\rho}\ddx{P}, \label{eq:x-momentum}\\
\ddt{v} + u\ddx{v} + v\ddy{v} + 2 \Omega u &=& -\frac{1}{\rho}\ddy{P}, \label{eq:y-momentum}\\
\ddx{u} + \ddy{v} &=& 0\label{eq:incompressibility},
\end{eqnarray}
where $x = r - r_0$ and $y$ define the local coordinate system. We
define the x and y velocities as $u$ and $v$, respectively, and the
gas pressure and density as $P$ and $\rho$.  Equations
(\ref{eq:continuity}), (\ref{eq:x-momentum}), (\ref{eq:y-momentum}),
(\ref{eq:incompressibility}) are the continuity equation, x and y
momentum equations, and condition of incompressibility, respectively.
The following solution to equations (\ref{eq:continuity}) -
(\ref{eq:incompressibility})
\begin{eqnarray}
u_0 &=& 0, \\
v_0 &=& -\frac 3 2 \Omega x,
\end{eqnarray}
defines the local background shearing flow.

We now discuss the \cite{Kida1981} and GNG solutions to equations
(\ref{eq:continuity}) - (\ref{eq:incompressibility}).  Both of these
solutions have the form,
\begin{eqnarray}\label{eq:steady state u}
u &=& \omega \chi^{-1} y, \\
v &=& -\omega \chi x.
\label{eq:steady state v}
\end{eqnarray}
Solutions that follow equation (\ref{eq:steady state u}) and
(\ref{eq:steady state v}) uniformly rotate on ellipses with
ellipticity $\chi >= 1$ at an angular frequency, $\omega$.  For
vortices in protoplanetary disks, $\omega$ and $\Omega$ have opposite
signs: the vortices are anticyclonic.  We note that while
the Kida solution was originally derived for purely shearing flows in
a fixed frame, i.e., for $\Omega = 0$ and without the tidal term,
$3\Omega^2 x$, it has been applied to the study of vortices in
protoplanetary disks as a means of collecting dust
\citep{Chavanis2000} and the overall stability of vortices to 3-d
effect \citep{Lesur2009}.

We solve for the equilibrium pressure distribution for the vortex
solutions given by equations (\ref{eq:steady state u}) and
(\ref{eq:steady state v}) by solving the steady state momentum
equations (\ref{eq:x-momentum}) and (\ref{eq:y-momentum}).  This gives
\begin{eqnarray}
\frac 1 {\rho} \ddx{P} &=& \left(\omega^2 + 3\Omega^2 - 2\Omega\omega\chi\right)x, \label{eq:P-xequil}\\
\frac 1 {\rho} \ddy{P} &=& \left(\omega^2 - 2\Omega\omega\chi^{-1}\right)y. \label{eq:P-yequil}
\end{eqnarray}

It is helpful to consider the pressure distribution in a coordinate
system better suited to these vortices.  As the steady state solution
for both the Kida and GNG vortices have elliptical streamlines with
ellipticity, $\chi$, we chose a coordinate system $(b,\phi)$ of the
form
\begin{eqnarray}
x &=& b\cos\phi,\label{eq:elliptical coord 1}\\
y &=& b\chi\sin\phi,  \label{eq:elliptical coord 2}
\end{eqnarray}
where $b$ is the semi-minor axis that characterizes a particular
ellipse and $\phi$ defines a position along that ellipse.  Note that
we have chosen our axes such that the y direction is the along major
axis of the ellipse. In this coordinate system,
\begin{eqnarray}
\ddx{} &=& \cos\phi\ddb{} - \sin\phi  \frac 1 b \ddphi{},\\
\ddy{} &=& \chi^{-1}\left(\sin\phi\ddb{} + \cos\phi \frac {1} b \ddphi{}\right).
\end{eqnarray}
Using the above and equations (\ref{eq:P-xequil}) and
(\ref{eq:P-yequil}), we find that
\begin{eqnarray}
\frac 1 {\rho} \ddb{P} &=& b\left(\omega^2 + 3\Omega^2\right)\cos^2\phi + b\chi^2\omega^2\sin^2\phi- 2b\Omega\omega\chi, \label{eq:P-bequil}\\
\frac 1 {b\rho} \ddphi{P} &=& \sin\phi\cos\phi\left[ \omega^2\left(\chi^2-1\right) - 3\Omega^2\right]. \label{eq:P-phiequil}
\end{eqnarray}

\subsection{The GNG Solution}\label{sec:gng}

GNG presented a solution for closed streamlines of the form given by
equations (\ref{eq:steady state u}) and (\ref{eq:steady state v}).  As
this is simpler than the Kida solution (discussed below), we focus on
this solution.  Assuming a polytropic relation between the pressure,
$P$ and $\rho$, we find a relation between $\omega$, $\Omega$, and
$\chi$ (GNG)
\begin{equation}\label{eq:gng}
\omega = \Omega \sqrt{\frac{3}{\chi^2-1}}.
\end{equation}
Applying the results of the above analysis of the pressure equilibrium
(eq. [\ref{eq:P-xequil}], [\ref{eq:P-yequil}], [\ref{eq:P-bequil}],
and [\ref{eq:P-phiequil}]) to the GNG vortex, we find:
\begin{eqnarray} \label{eq:dpdb gng}
\frac 1 {\rho} \ddb{P} &=& b\Omega^2\left(\frac {3\chi^2}{\chi^2 - 1} - 2\sqrt{\frac{3\chi^2}{\chi^2 - 1}}\right),\\
\frac 1 {b\rho} \ddphi{P} &=& 0.
\end{eqnarray}
The pressure distribution of the GNG vortex is very simple compared to
the Kida case.  Its pressure gradient is zero along $\phi$, and the
pressure gradient is constant between streamlines.  Note, however, the
pressure gradient in x-y coordinates is not constant moving along a
streamline due to their ellipticity.  The pressure is negative outward
(high pressure center) for $\chi > 2$ and inward (low pressure center)
for $\chi < 2$. Note that for $\chi$ = 2, we find $\omega = -\Omega$,
which describes epicyclic motion and that the pressure gradient is
zero, i.e., epicyclic motion demands no additional forces.

\subsection{The Kida Solution}\label{sec:kida}

\cite{Kida1981} (see also \citealt{Chavanis2000,Lesur2009}) presented
an exact solution to the 2-D Euler equations
(eqs.[\ref{eq:continuity}] - [\ref{eq:incompressibility}] with $\Omega
= 0$) for a background shear and an elliptic patch
with uniform vorticity, $\vorticity$.  In the core, the vortex
streamlines follow equations (\ref{eq:steady state u}) and
(\ref{eq:steady state v}), while outside of the core, the streamlines
asymptotically map onto the background shearing flow.  \cite{Kida1981}
showed (see \citealt{Chavanis2000}, his Appendix A) that $\vorticity$,
$\chi$, and $\Omega$ for a time steady vortex is given by
\begin{equation}
\frac {3\Omega}{2\vorticity} = \frac {\chi(\chi - 1)}{1 + \chi^2}.  
\end{equation}
Fluid elements in these ellipses move at a constant angular velocity,
\begin{equation}\label{eq:kida}
\omega = \frac{\vorticity\chi}{1 + \chi^2} = -\frac{3\Omega}{2(\chi - 1)}.
\end{equation}
We should note that the \citet{Kida1981} solution to the
incompressible 2-D Euler equations enforces a nontrivial pressure
distribution (see also \citealt{Lesur2009}).  Applying the same
results of the above analysis of the pressure equilibrium
(eq. [\ref{eq:P-xequil}], [\ref{eq:P-yequil}], [\ref{eq:P-bequil}],
and [\ref{eq:P-phiequil}]) to the Kida vortex as we have done for the
GNG vortex, we find:
\begin{eqnarray}
\frac 1 {\rho} \ddb{P} &=& \frac{3b\Omega^2}{4\left(\chi - 1\right)^2}\left[\left(4\chi^2 - 8\chi + 7\right)\cos^2\phi + 3\chi^2\sin^2\phi- 4\chi\left(\chi - 1\right)\right],\\
\frac 1 {b\rho} \ddphi{P} &=& -\frac {3\Omega^2}{4\left(\chi - 1\right)^2}\left(\chi^2 - 8\chi + 7\right)  \sin\phi\cos\phi.
\end{eqnarray}
The pressure distribution of the Kida vortex is rather complicated:
the pressure gradient in the $\phi$ direction is nontrivial, and for
$\chi > 4$, the radial pressure gradient can vary from positive outward
to negative outward where $\sin\phi = 1$ (the long axis).

\section{Stability of Protoplanetary Vortices}\label{sec:stability}

In general, spatially varying flows are not amenable to the WKB-type
analysis (see the discussion in \citealt{Bayly1988}).  In the
appendix, \S\ref{sec:simple}, we discuss a simple case of a terrestial
vortex, where the base flow is axisymmetric and therefore allows us
apply a cylindrical coordinate system. With this transformation the
flow is simple and the perturbation equations separable. For spatially
varying flows such elliptical vortices, this is not possible and
different techniques have to be brought to bear.

One very powerful technique developed by \cite{Lifschitz1991} combines
Floquet analysis with short wavelength WKB analysis. It is suitable
for analyzing perturbations that grow both exponentially and
algebraically in time or spatially varying flows.  We provide a brief
summary of this technique below and utilize it to analyze the stability of
the equilibrium vortex solutions of GNG and \cite{Kida1981}.

\subsection{Floquet Theory}\label{sec:floquet}

Here, we follow the logic of \cite{Lifschitz1991} and \cite{Sipp2000}.  We
begin by perturbing inviscid incompressible Euler equations on the
shearing sheet (eqs.[\ref{eq:continuity}] -
[\ref{eq:incompressibility}]) to find
\begin{eqnarray}
\DDt{\delta\rho} + \delta\uvec\cdot\nablavec\rho &=& 0,\label{eq:pert_continuity} \\
\nablavec\cdot\delta\uvec &=& 0, \label{eq:pert_incompressibility} \\
\DDt{\delta\uvec} + \delta\uvec\cdot\nablavec\uvec + 2\Omegavec\times\delta\uvec &=& -\frac 1 {\rho}\nablavec{\delta P} + \frac {\delta\rho}{\rho}\frac 1 {\rho}\nablavec{P},\label{eq:pert_momentum}
\end{eqnarray}
where we have written the momentum equation in vectorial format, and
$d/dt = \partial/\partial t + \uvec\cdot\nablavec$.  Now we will take
perturbations of the form:
\begin{equation}
  \left( 
  \begin{array}{c}
    \delta\uvec \\ \delta P \\ \delta \rho 
    \end{array}
  \right)
= \exp\left(\frac{i\Phi(\xvec,t)}{\epsilon}\right)
  \left[
    \left( 
    \begin{array}{c}
      \utildevec \\ \Ptilde \\ \rhotilde 
    \end{array}
    \right)(\xvec,t) + 
    \epsilon\left( 
    \begin{array}{c}
      \utildeveceps \\ \Ptildeeps \\ \rhotildeeps 
    \end{array}
    \right)(\xvec, t)
    \right], 
\end{equation}
where $\Phi$ is a real phase function and $\epsilon$ is a small
parameter. Inserting this {\it ansatz} into equation
(\ref{eq:pert_incompressibility}) gives
\begin{eqnarray}
\nablavec\Phi\cdot\utildevec = 0, \label{eq:kdotu}\\ 
\nablavec\cdot\utildevec = 0,
\end{eqnarray}
to the lowest order and the next order in $\epsilon$, respectively.
It is helpful to define the local wavevector, $\kvec \equiv \nablavec\Phi$.
Inserting the same {\it ansatz} into equation(\ref{eq:pert_continuity}) gives
\begin{eqnarray}
\DDt\Phi = 0, \label{eq:phase}\\ 
\DDt{\rhotilde} + \utildevec\cdot\nablavec\rho = 0.
\end{eqnarray}
The phase function $\Phi$ is conserved along a streamline. 
Finally, the perturbed momentum (\ref{eq:pert_momentum}) gives 
\begin{equation}\label{eq:momentum_tilde}
\DDt{\utildevec} + \utildevec\cdot\nablavec\uvec + 2\Omegavec\times\utildevec=
-\frac {i}{\rho} \kvec\cdot\Ptildeeps + \frac{\rhotilde}{\rho}\frac 1 {\rho}\nablavec P.
\end{equation}
Note that in deriving \ref{eq:momentum_tilde} we use the fact that to lowest order
($1/\epsilon$), the perturbed pressure gradient term in \ref{eq:pert_momentum} gives
\begin{equation}
  \frac{i \kvec \Ptilde}{\rho} = 0,
\end{equation}
implying that $\Ptilde$ and thus $\nablavec \Ptilde$ are zero everywhere,
as the local wavevector is in general not zero.

Now we project equation (\ref{eq:momentum_tilde}) onto the line
perpendicular to $\kvec$ to eliminate $\Ptildeeps$ using the operator
$I - \kvec\kvec/k^2$, where $I$ is the identity.  This yields
\begin{equation}\label{eq:floquet_momentum}
\DDt\utildevec = \left(2\frac{\kvec\kvec} {k^{2}} - I\right) \cdot\Lmatrix\cdot\utildevec + 
\left(\frac{\kvec\kvec} {k^{2}} - I\right) \cdot 2\Omegavec\times\utildevec - \frac {\rhotilde}{\rho^2}\left(\frac{\kvec\kvec} {k^{2}} - I\right)\cdot\nablavec P,
\end{equation}
where the velocity gradient tensor $\Lmatrix$ is defined as 
\begin{equation}
\Lmatrix = \nablavec\uvec.
\end{equation}
The other parts of the equation of motion are 
\begin{eqnarray}
\DDt{\rhotilde} + \utildevec\cdot\nablavec\rho &=& 0, \label{eq:floquet_rho}\\
\DDt{\xvec} &=& \uvec, \label{eq:floquet_x}\\
\DDt{\kvec} &=& -\Lmatrix^{T} \kvec \label{eq:floquet_k},
\end{eqnarray}
where we derived the last equation by applying $\nablavec$ to equation 
(\ref{eq:phase}).  Equations
(\ref{eq:floquet_momentum}) and (\ref{eq:floquet_rho}) -
(\ref{eq:floquet_k}) are the evolution equations for the
position ($\xvec$), perturbed density ($\rhotilde$), perturbed
velocity ($\utildevec$) and perturbation wavevector ($\kvec$) of the
fluid perturbation.  As $\xvec$ now has an explicit time dependence,
this system of equations is amenable to Floquet analysis.

\subsection{Instability Analysis}\label{sec:instability}
\begin{figure}
  \plotone{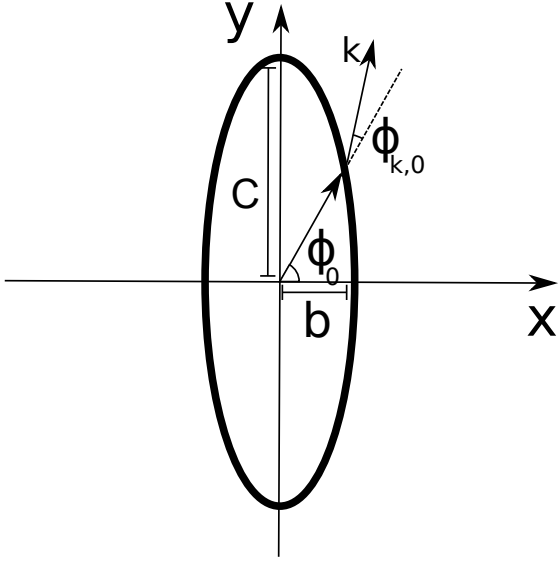}
  \caption{The basic vortex geometry. A sample streamline is in bold,
    with semi-minor axis b and semi-major axis c. The ellipticity
    $\chi = c/b  = 4$ in this case. $\phi_0$ shows the initial angle
    of the radius vector with the $x$ axis while $\phi_{k,0}$ is the
    initial angle of the perturbation wavevector $\mathbf{k}$ with
    respect to the radius vector.}
  \label{f:k_angle}
\end{figure}
Taking the equilibrium solution (eq.[\ref{eq:steady state u}] and
[\ref{eq:steady state v}]) posed above, we find
\begin{equation}\label{eq:Lmatrix}
\Lmatrix = \omega \left(\begin{array}{cc}
  0 & \chi^{-1} \\ -\chi & 0
  \end{array}\right).
\end{equation}
Using this result to solve equations (\ref{eq:floquet_x}) and
(\ref{eq:floquet_k}), we find
\begin{eqnarray}
x &=& b\cos(\omega t + \phi_0), \\
y &=& -\chi b\sin(\omega t + \phi_0), \\
k_x &=& k_0\cos(\omega t + \phi_{k,0}), \\
k_y &=& -\chi^{-1} k_0\sin(\omega t + \phi_{k,0}),
\end{eqnarray}
where $\phi_0$ and $\phi_{k,0}$ are the initial phase of the
coordinate and the wavevector and $b$ is the semi-minor axis of the
elliptical streamline (see figure~\ref{f:k_angle}).  We can also use the result
$\kvec\cdot\utildevec = 0$ (eq. [\ref{eq:kdotu}]) to write
\begin{eqnarray}
\tilde{u}_x &=& a(b,t)  \sin(\omega t + \phi_{k,0}), \label{eq:utildex}\\
\tilde{u}_y &=& a(b,t)  \chi\cos(\omega t + \phi_{k,0}),\label{eq:utildey}
\end{eqnarray}
where $a(b,t)$ determines the overall normalization of $\utildevec$.
Applying the results of equations (\ref{eq:utildex}) -
(\ref{eq:utildey}) to equations (\ref{eq:floquet_momentum}) and
(\ref{eq:floquet_rho}), we find
\begin{eqnarray}
\DDt{a} &=& 2\Lambda^{-1}\omega\left(\chi^2 - 1\right)\cos(\omega t +
  \phi_{k,0})\sin(\omega t + \phi_{k,0})a \nonumber \\
&&+ \Lambda^{-1}\left[ \left(\omega^2 + 3\Omega^2\right)\cos(\omega t + \phi_{0})\sin(\omega t + \phi_{k,0}) \right.\nonumber \\
&&- \chi^2\omega^2 \sin(\omega t + \phi_{0})\cos(\omega t + \phi_{k,0})
\nonumber\\
&& \left.- 2\Omega\omega\chi\sin(\phi_{k,0} - \phi_0)
\right]b\frac{\rhotilde}{\rho}, \label{eq:a(t)}\\
\DDt{\rhotilde} &=& -a \sin(\phi_{k,0} - \phi_{0})\ddb\rho, 
\label{eq:rhotilde(t)}
\end{eqnarray}
where $\Lambda = \chi^2\cos^2(\omega t + \phi_{k,0}) + \sin^2(\omega t
+ \phi_{k,0})$.  Without loss of generality, we can choose $\phi_0 =
0$.  Thus, equations (\ref{eq:a(t)}) and (\ref{eq:rhotilde(t)}) depend
only on the initial angle of the wavevector, $\phi_{k,0}$. Equation
(\ref{eq:a(t)}) explicitly depends on the semi-minor axis
$b$. However, because $b$ is independent of time and appears only in
the $\tilde{\rho}$ term in combination with $\rho$, we can eliminate
it by absorbing it into the background density gradient, replacing
$\partial \rho/\partial b$ with $\partial \ln \rho/\partial \ln b$ in equation
(\ref{eq:rhotilde(t)}).

The first term on the RHS of equation (\ref{eq:a(t)}) results from the
first two terms of equation (\ref{eq:floquet_momentum}).  The second
term results from the density gradient within the vortex.  For a zero
density gradient, the integral over a period $T = 2\pi\omega^{-1}$
of equation (\ref{eq:a(t)}) is zero as this first term on the RHS is
an odd function.  Hence, in the absence of a density gradient, no
exponentially growing modes exists in two dimensions.  In addition,
for an initially aligned wavevector (i.e. $\phi_{k,0} = 0$) and
$\rhotilde(t=0) = 0$, there is also no growth, regardless of the
presence of a density gradient.

Equations (\ref{eq:rhotilde(t)}) and (\ref{eq:a(t)}) constitute a
system of linear ODEs, which depends on initial conditions for $a$ and
$\rhotilde$, the initial angle of the wavevector $\phi_{k,0}$, the
background density profile, $\partial\rho/\partial b$, and the vortex
ellipticity, $\chi$.  However, as we are interested in the asymptotic
behavior of perturbations, i.e., growth or no growth, we are
interested in the parameter range of $\phi_{k,0}$, $\chi$, and
$\partial\rho/\partial b$, which yield stability or instability.  To
this end, we define a growth rate $\gamma$ where $\rhotilde(t=\tau) =
e^{\gamma \tau} f(\tau)$ and $f(t)$ is a period function over the
vortex rotation period, $\tau = 2\pi/\omega$.  To find $\gamma$, we
follow the general procedure of Floquet analysis on equations
(\ref{eq:a(t)}) and (\ref{eq:rhotilde(t)}).  For linearly independent
initial conditions, e.g., $(a(t=0)=1,\rhotilde(t=0)=0)$ and
$(a(t=0)=0,\rhotilde(t=0)=1)$,\footnote{Note that we are ultimately interested in the growth factor over the period $\tau$, i.e., $e^{\gamma\tau} = \rhotilde(\tau)/\rhotilde(0)$ or $=a(\tau)/a(0)$.  Hence, it makes sense to rescale this linear problem in terms of the initial perturbation, i.e., $a(t=0) = 1$ or $\rhotilde(t=0) = 1$.} we can integrate equations
(\ref{eq:a(t)}) and (\ref{eq:rhotilde(t)}) over one period
$2\pi/\omega$ and solve for the most unstable eigenvalue of the
resulting matrix \citep[e.g.][]{BenderOrszag,Lesur2009}.

\begin{figure}
  \plotone{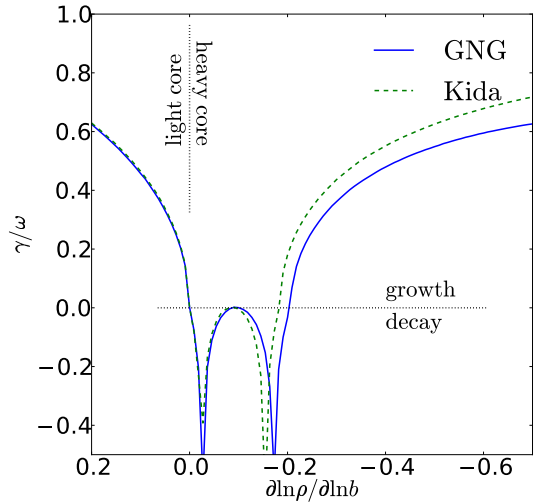}
  \caption{Stability of protoplanetary vortices as a function of the
    density contrast.  For any vortex that has a light core, i.e.,
    $d\ln\rho/d\ln b > 0$, we find a purely growing instability, i.e.,
    the Vortical Rayleigh-Taylor Instability, which we discuss in
    \S\ref{sec:light cores}.  For sufficiently heavy cores, there is
    an instability, which we discuss in \S\ref{sec:heavy cores}.}
  \label{f:density_growth_rate}
\end{figure}
\begin{figure}
  \plotone{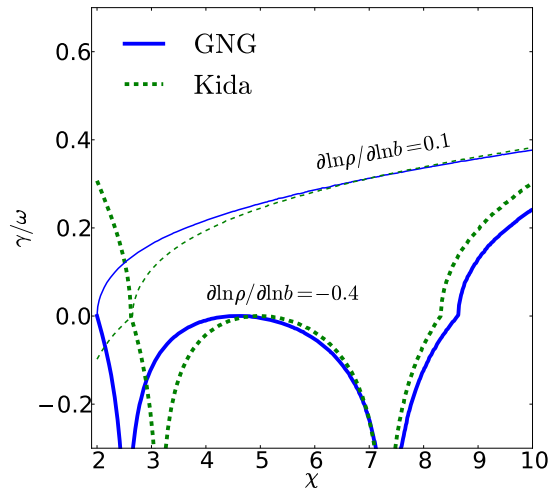}
  \caption{Stability of protoplanetary vortices as a function of the
    $\chi$.  For any vortex that has a light core, i.e.,
    $d\ln\rho/d\ln b > 0$, we find a purely growing instability for
    any value of $\chi > 2$. For sufficiently heavy cores, i.e., the
    $\partial\ln\rho/\partial\ln b = -0.4$ case, Instability also
    demands a sufficiently large $\chi$.}
  \label{f:chi_growth_rate}
\end{figure}
\begin{figure}
  \includegraphics[width=0.95\columnwidth]{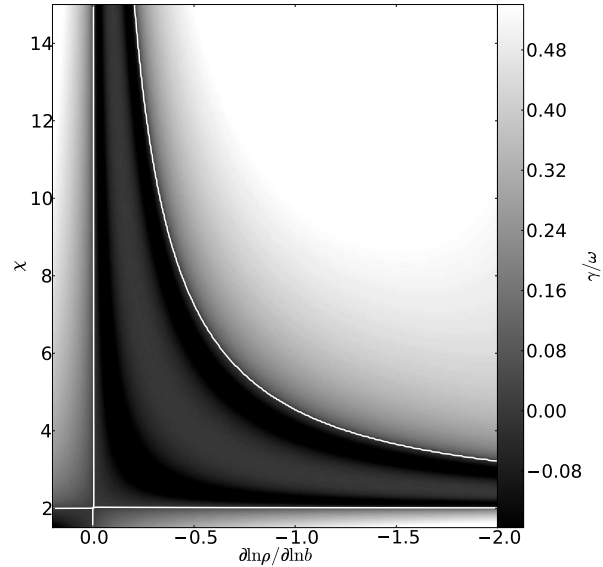}
  \caption{Growth rates as a function of $\chi$ and $\partial \ln
    \rho/\partial \ln b$ for the GNG vortex. The heavy white contour
    denotes the stability boundary ($\gamma/\omega = 0$). }
  \label{f:image_growth}
\end{figure}

In Figure \ref{f:density_growth_rate} we show the effect of different
density contrasts on the growth rate for fixed $\chi = 10$.  Here we
consider both light cores ($\partial\ln\rho/\partial\ln b > 0$) and
heavy cores ($\partial\ln\rho/\partial\ln b < 0$).  There are two
regions of instability: one for light cores and and one for
sufficiently heavy cores, which we discuss below.  We show the effect
of the vortex ellipticity $\chi$ on $\gamma$ in Figure
\ref{f:chi_growth_rate} for the GNG (solid lines) and the Kida (dashed
lines) for the light cores (light lines) and heavy cores (heavy
lines).
For the light core case, we find that as $\chi$ increases, the growth
rate in terms of $\omega$ also increases.  However, there is generally
always growth (with the exception of the low $\chi$ Kida case). In the
heavy core case, there is no growth until $\chi$ is sufficiently large
and the core sufficiently dense compared to the ambient flow.  We
summarized the growth rate for $\phi_{k,0} = \pi/2$ as a function of
both $\chi$ and $\partial\ln\rho/\partial\ln b$ in Figure
(\ref{f:image_growth}).  In the case of the light core,
$\partial\ln\rho/\partial\ln b > 0$, we find instability for any
density contrast.  The heavy core case is more complicated.  For small
$\partial\ln\rho/\partial\ln b$ and small $\chi$, there is no
instability. However, once $\partial\ln\rho/\partial\ln b$ and $\chi$
are sufficiently large, growth sets in.  Growth can occur for smaller
$\partial\ln\rho/\partial\ln b > 0$ if $\chi$ is larger.  In between
the two regions of growth, i.e., light cores and sufficiently heavy
cores, $\gamma < 0$, indicating damping.  The two heavy white lines
marks the region of neutral stability.  The physics of instability of
these two light and heavy core cases are somewhat different, which we
now discuss in \S\ref{sec:light cores} and \S\ref{sec:heavy cores}.

\subsection{Vortices with Light Cores}\label{sec:light cores}

The instability mechanism for light cores is analogous to the
Rayleigh-Taylor instability.  Recall that the protoplanetary vortices
are high pressure regions.  The pressure forces thus point outward
and must be balance by a combination of centripetal and centrifugal
forces which must point inward.  If we "unroll" this vortex, we see
that the combination of centripetal and centrifugal forces, which
oppose the pressure force, is analogous to gravity in a pressure
support atmosphere.  Hence, a light core in this context is equivalent
to making the material less dense where the pressure is largest, i.e.,
putting denser material on top of less dense material.  This is
subject to a Rayleigh-Taylor-like instability, which we call the
Vortical Rayleigh-Taylor Instability (VRTI).  In the appendix, we
present a simple example of this instability in a terrestial vortex to
make more precise the analogy between the Rayleigh-Taylor Instability
(RTI) and the VRTI.  We note, however, that in the terrestial example
presented in the appendix that the condition for instability is a
heavy core as opposed to a light core.  This results from the fact
that terrestial vortices are low pressure regions, whereas
protoplanetary vortices are high pressure regions.

We now demonstrate the behavior of perturbations for light cores by
plugging $\omega$ for GNG (or Kida) vortices (eq. [\ref{eq:gng}]) into
equation (\ref{eq:a(t)}), we integrate the evolution of equations
(\ref{eq:a(t)}) and (\ref{eq:rhotilde(t)}) over several periods.  We
note that equation (\ref{eq:a(t)}) admits a purely analytic solution
when $\rhotilde = 0$.  Integrating both sides with this in mind, we
find
\begin{equation}\label{eq:analytic}
a(t) = a_0\frac{\chi^2 + (1-\chi^2)\sin^2\left(\phi_{k,0}\right)}{\chi^2 + (1-\chi^2)\sin^2\left(\omega t + \phi_{k,0}\right)},
\end{equation}
where $a_0$ is fixed by initial conditions.  The analytic solution
(\ref{eq:analytic}) represents the evolution of a perturbation that is
purely advected along in the flow.  More complex cases are solved
numerically.

\begin{figure*}
  \plottwo{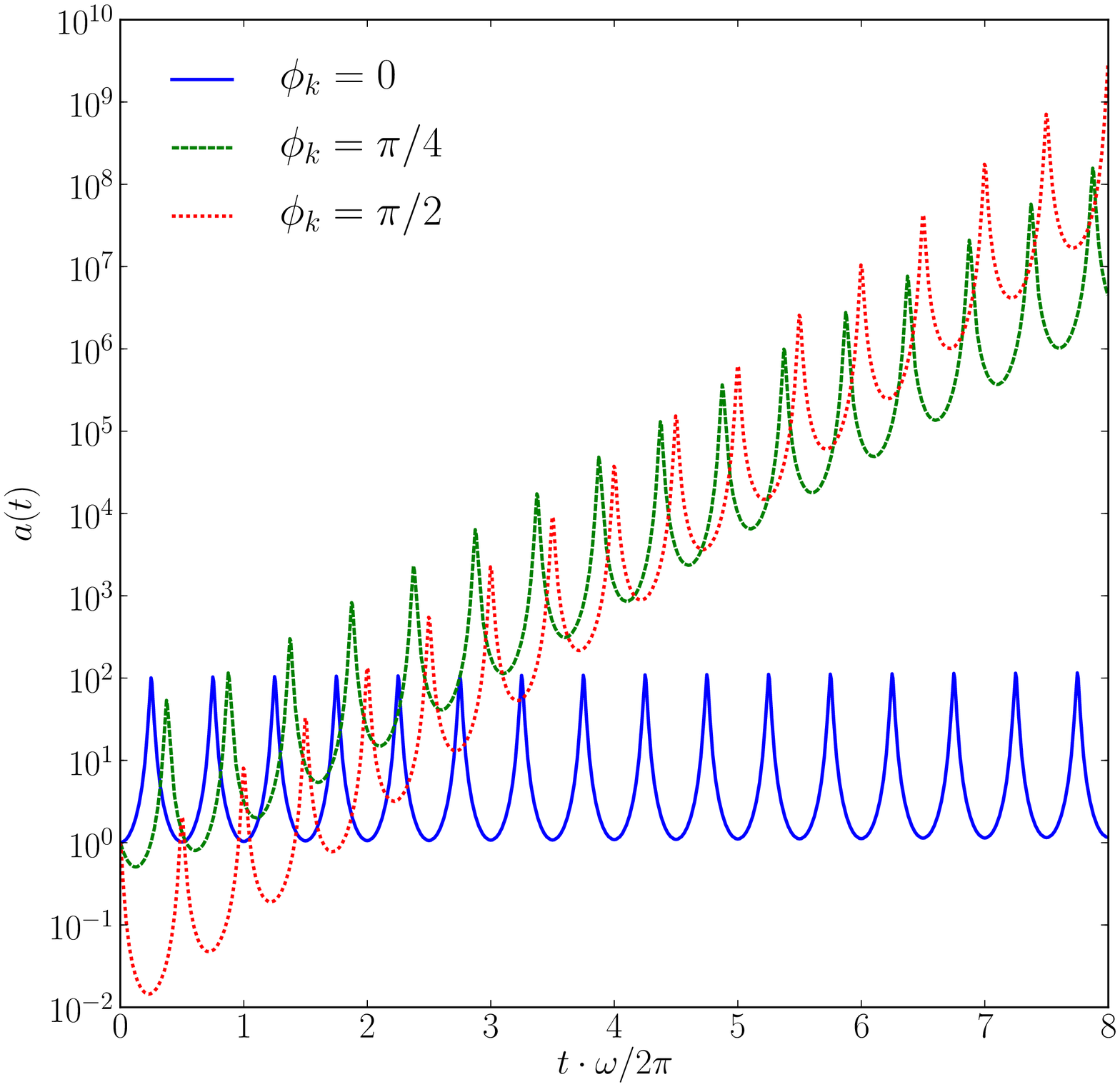}{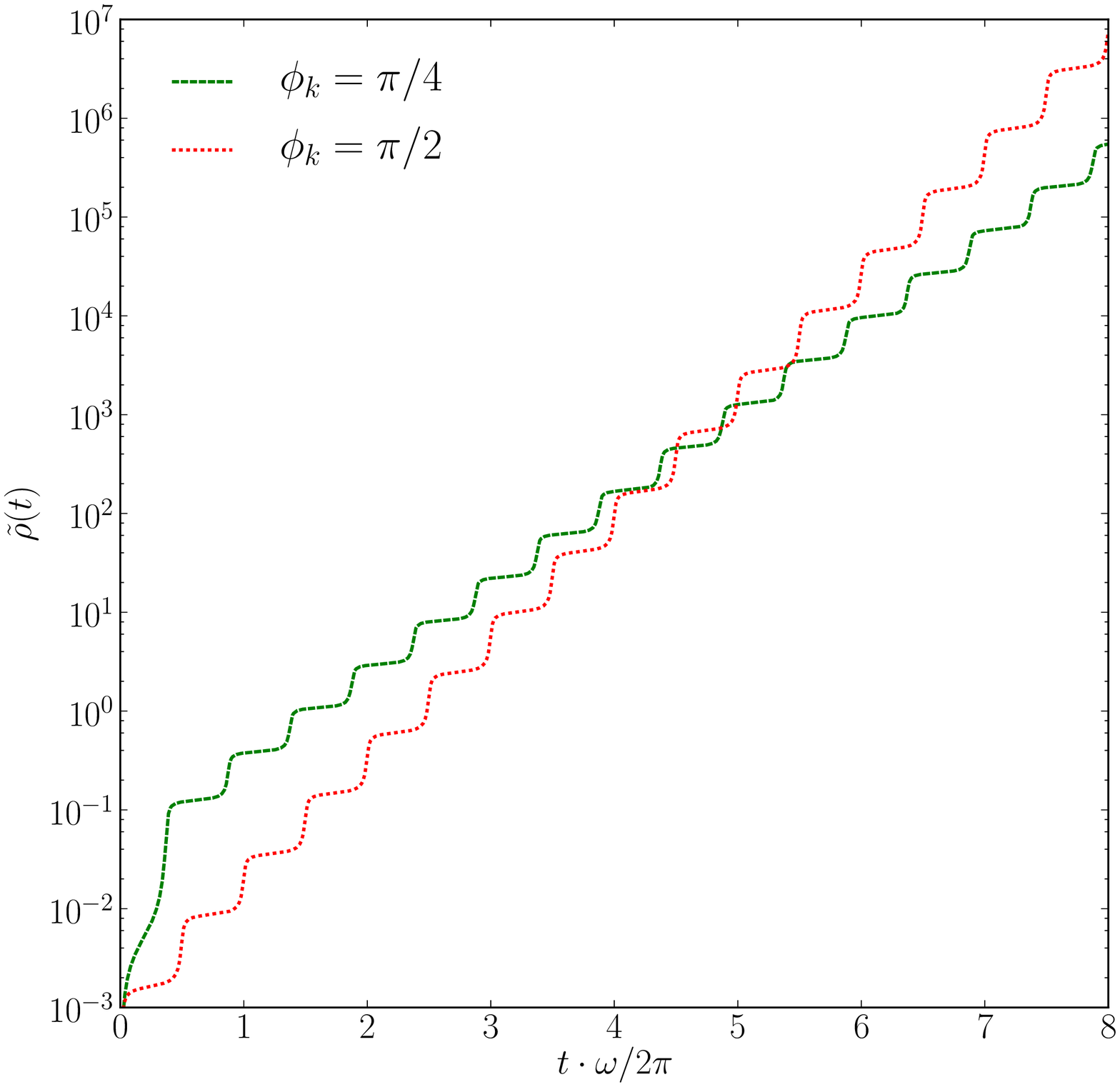}

  \caption{Velocity (\textit{left}) and density (\textit{right})
    perturbations as a function of time for three initial wavevector
    orientation angles $\phi_{k,0}$ for light cores
    ($\partial\ln\rho/\partial\ln b = 0.01$). Note that $\phi_k = 0$
    has been dropped from the right plot because it has zero
    growth. This case is show on the left plot for illustrativ
    purposes. There is no growth for the $\phi_{k,0} = 0$ case, and
    the growth rate is maximum for for $\phi_{k,0} = \pi/2$.}
  \label{f:phik_gng}
\end{figure*}

Figure \ref{f:phik_gng} shows the behavior of $a(t)$ and
$\tilde{\rho}(t)$ for the initial conditions $a(t=0) = 1$ and
$\rhotilde(t=0) = 0$ and a background density profile of
$\partial\ln\rho/\partial\ln b = 0.01$, i.e., a light core.  The
background density profile corresponds to a one percent decrease in
the density across the vortex.  The velocity amplitude $a(t)$ for the
$\phi_{k,0} = 0$ case shows oscillatory behavior between 1 and 100,
but no long term growth occurs. Correspondingly, the density
perturbation remains zero for all time and is not shown in the right
panel. Indeed, the solution's behavior precisely follows the analytic
result of equation (\ref{eq:analytic}), verifying the accuracy of our
numerical integration.  On the other hand, both $\phi_{k,0} = \pi/4$
and $\phi_{k,0} = \pi/2$ show significant growth, and the asymptotic
growth rate, i.e., the slope of the trend, is maximized for $\pi/2$.
This is unsurprising given the discussion in the appendix and the
analogy with the RTI, where we found that growth rates are maximized
when the wavevector is parallel to vortex streamlines.

\begin{figure}
  \plotone{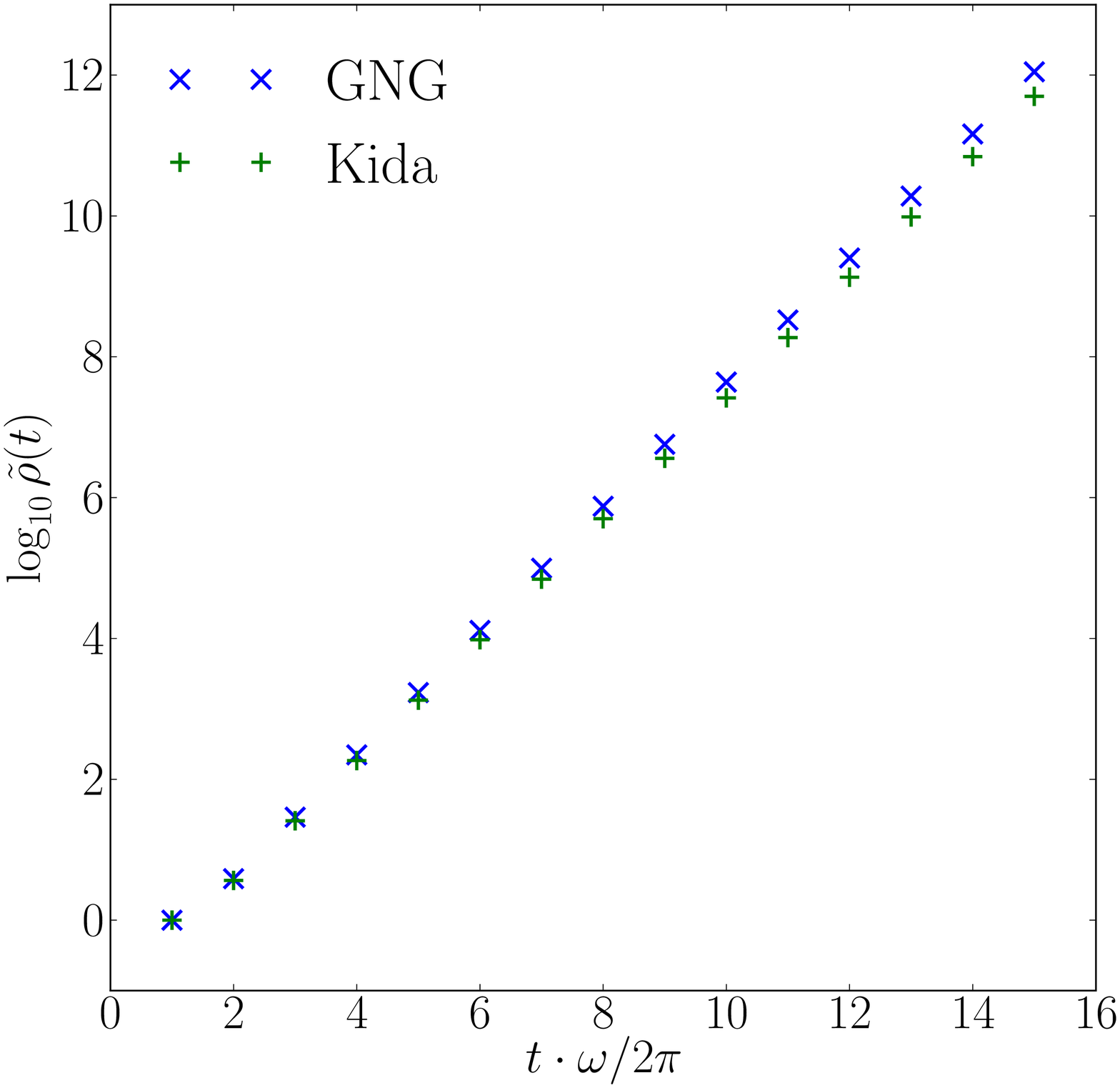}
  \caption{A comparison of the GNG and Kida background states for the
    $\phi_{k,0} = \pi/2$, $\chi = 5$ case for the light core case
    ($\partial\ln\rho/\partial\ln b = 0.01$). The figure shows density
    perturbation as a function of time for both background states.}
  \label{f:gng_kida_comparison}
\end{figure}

Similarly, we plug in $\omega$ for Kida vortices (eq. [\ref{eq:kida}])
into equation (\ref{eq:a(t)}) and integrate its evolution over several
periods.  Before comparing the behavior of the Kida vortices with that
of the GNG vortices, we first note that from Figure \ref{f:phik_gng},
the velocity and density shows both short timescale periodic
fluctuations, a result of the spatially inhomogeneous flow field, i.e.,
vortex, and long term behavior.  As we are only interested in long
term behavior, we sample $\rhotilde$ where $\omega t$ is an integer
multiple of $2\pi$ for both vortices and show the results in
Figure~\ref{f:gng_kida_comparison}.  The comparison clearly shows that
the background state makes no difference in the qualitative behavior
of the instability and little difference in the quantitative behavior
of the growth rate. 

\subsection{Vortices with Heavy Cores}\label{sec:heavy cores}

We now discuss the heavy core case.  To help elucidate the physics, we
first consider the effective gravity that counteract the pressure
forces in a GNG vortex, i.e., $\rho^{-1}\partial P/\partial x =
-g_x,\ \rho^{-1}\partial P/\partial y = -g_y $.  From equation
(\ref{eq:dpdb gng}), we know that the pressure drop between vortex
streamlines is constant.  Hence, the effective gravity is 
\begin{eqnarray} \label{eq:gx}
g_x &=& -\rho^{-1}\frac {\partial P}{\partial b} \frac{db}{dx} \\
g_y &=& -\rho^{-1}\frac {\partial P}{\partial b} \frac{db}{dy}, \label{eq:gy}
\end{eqnarray}
where $b = \sqrt{x^2 + y^2/\chi^2}$.  Hence computing the magnitude of
the effective gravity is, thus,
\begin{equation}
g = \sqrt{g_x^2 + g_y^2} = g_0\sqrt{\cos^2\phi + \frac {\sin^2\phi}{\chi^2}},
\end{equation}
where 
\begin{equation}
g_0 = b\Omega^2\left(\frac{3\chi^2}{\chi^2 - 1} -
2\sqrt{\frac{3\chi^2}{\chi^2 - 1}}\right) = {\rm const}.
\end{equation}

\begin{figure}
  \plotone{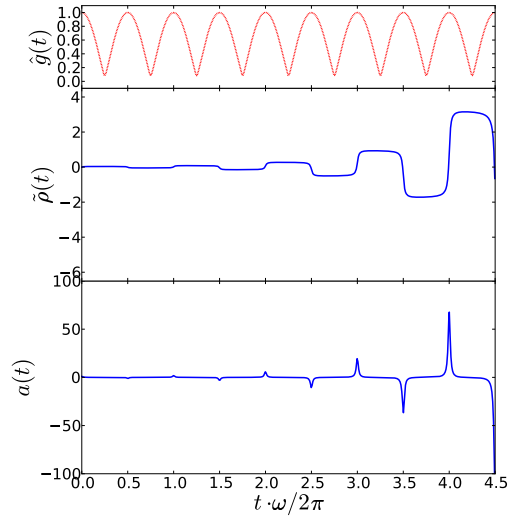}
  \caption{The effective gravitational acceleration $\hat{g}(t) =
    g(t)/g_0$ normalized to the maximum gravitational
    acceleration, perturbed density (in arbitrary units), and $a(t)$
    (again in arbitary units). Note that the peaks of $g(t)$
    correspond to the peaks in $a(t)$ and changes in $\rhotilde(t)$,
    which correspond to periodic kicks at roughly half the rotation
    period of the vortex.}
  \label{f:phik_gng_g}
\end{figure}

We now plot the behavior of the effective gravity ($\hat{g}(t) = g(t)/g_0$) and the
growing perturbations ($a(t)$ and $\tilde{\rho}(t)$) in
Figure \ref{f:phik_gng_g} for $\phi_{k,0} = \pi/2$.  Note that
perturbed density ($\tilde{\rho}(t)$) and velocity ($a(t)$) changes
only at intervals where $g(t)$ is large.  This is unsurprising for
large $\chi$, the effective gravity will vary between $g_0/\chi$ and
$g_0$.  These peaks in the effective gravity is reached for $y=0$,
i.e., on the minor axis of the elliptical vortex.  This is reasonable
from a physical perspective as it is here that the distance between
vortex streamlines is minimal while the pressure change between vortex
streamlines remain unchanged (for GNG vortices). Thus, the force is
maximal there.
This effective gravity, which is time dependent (as fluid is advected
along a vortex), is akin to a kick.  The relative strength of these
kicks depend on $\chi$ and the period of these kicks is exactly half
of the rotation period of the vortex as a fluid element moving along a
streamline crosses the minor axis twice a rotation period.  As the
modes we are following is akin to radial gravity modes, the density
contrast determines their period. Hence, the minimum density contrast
required for instability has an obvious intepretation: the
radial gravity mode must also have a period that is comparable to the
rotation period of the vortex to it to couple successfully to the kick
and develop overstable oscillations as the plot of $\rhotilde$ in
Figure \ref{f:phik_gng_g} shows. In addition, the minimum $\chi$
needed for instability as demonstrated in Figure
\ref{f:chi_growth_rate} and \ref{f:image_growth} results from
requiring each kick to be sufficiently strong.

\begin{figure}
  \plotone{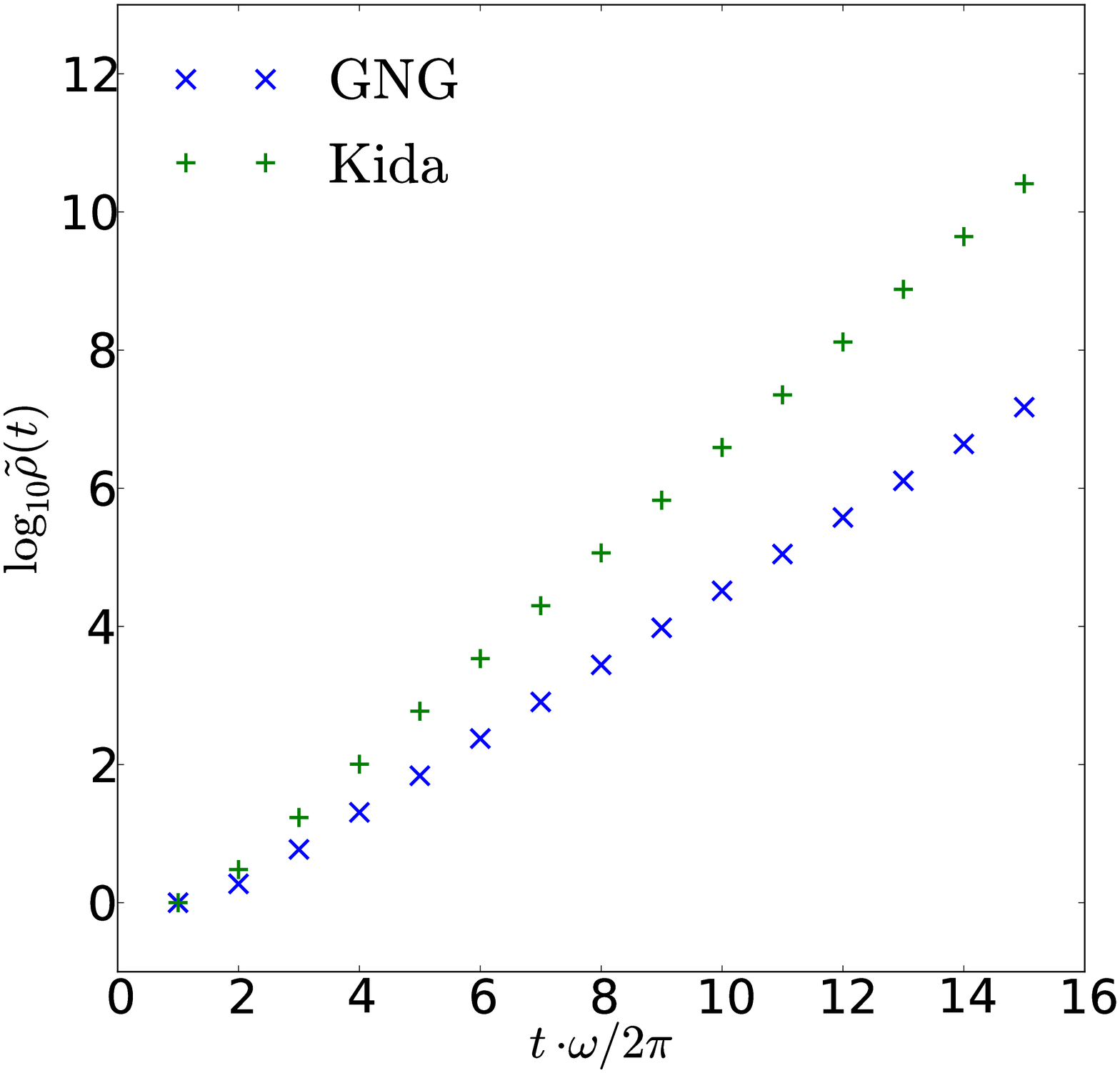}
  \caption{Same as Figure \ref{f:gng_kida_comparison} except for the
    heavy core case.  Here $\chi=12$ and $\partial\ln\rho/\partial\ln
    b = -0.3$.}
  \label{f:gng_kida_heavy_comparison}
\end{figure}

Similarly, we can make the same comparison between the GNG and Kida
vortex in the heavy core case in Figure
\ref{f:gng_kida_heavy_comparison} as we have done in the light core
case in Figure \ref{f:gng_kida_comparison}. Again, we plug in $\omega$
for Kida vortices (eq. [\ref{eq:kida}]) into equation (\ref{eq:a(t)})
and integrate its evolution over several periods but this time for a
background density gradient of $\partial\ln\rho/\partial\ln b = -0.3$
and $\chi = 12$.  Again we discard short timescale periodic
fluctuations and sample $\rhotilde$ where $\omega t$ is an integer
multiple of $2\pi$ for both vortices.  The comparison shows that the
growth rate of the HCI is more dependent on the vortex solution (Kida
vs. GNG) than the VRTI.  Whereas Figure \ref{f:gng_kida_comparison}
shows that the amplitude of the density perturbation for the Kida and
GNG vortices track each other fairly closely, these amplitude diverge
much more strongly in the HCI as shown in Figure
\ref{f:gng_kida_heavy_comparison}.  

\section{Discussion}\label{sec:discussion}

The analysis of the preceding section demonstrates that vortices with
light cores or sufficiently heavy cores are unstable to the VRTI and
HCI respectively.  Figures \ref{f:density_growth_rate} and
\ref{f:image_growth} shows that growth occurs on a few vortex rotation
periods.  Moreover, these instability appears to be robust and its
detailed physics are independent of the vortex model used (either GNG
or Kida). Having demonstrated the basic physics of these
instabilities, we turn now to its application to planetesimal
formation.  We will first review some of the physics of dust trapping
in vortices.

If planetesimals form by gravitational collapse and fragmentation of a
dust sublayer \citep{Safronov1969,Goldreich1973}, then this layer must
have a Toomre, $Q < 1$ (but also see \citealt{Ward2000}), which
implies that the velocity dispersion of the sublayer must be below:
\begin{equation}\label{eq:Q-criterion}
\sigma_{\rm d} < \frac {\pi G\Sigma_{\rm d}}{2\Omega} \approx 5 \,{\rm cm\,s}^{-1}
\end{equation}
for the minimal mass solar nebula (MMSN), where $\Sigma_{\rm d}$ is
the surface density of the dust layer.  For a laminar disk, this
criterion is amply fulfilled if the dust is allowed to settle to the
midplane.  However, as the dust collects near the midplane, it is
subject the induced Kelvin-Helmholtz instabilities with the overlying
gas layers
\citep{Weidenschilling1993,Cuzzi1993,Chiang2008,Barranco2009}.
Therefore, the dispersion of the dust layer is closer to a few ${\rm
  m\,s}^{-1}$.  Hence the surface density must be enhanced by a factor
of $\sim 20-100$ (effectively the $Q$ of the gaseous disk) so that
gravitational instability can operate \citep{Chavanis2000}.

More careful considerations suggest this enhancement of $\sim 20-100$
may be a severe overestimate and that only enhancement of order a few is
needed in the high metallicity disks which preferentially form planets \citep{Youdin2002,Johansen2009}.  In any case, vortices are one avenue by such a dust
surface density enhancement can be achieved as \cite{Barge1995} first
pointed out.  The timescale for dust to connect and concentrate in
vortices, i.,e., the capture timescale, $t_{\rm capt}$, can be fairly
rapid, i.e., $t_{\rm capt} \sim t_{\rm dyn}$, when $t_{\rm stop} \sim
t_{\rm dyn}$ as pointed out by
\cite{Barge1995,Tanga1996,Chavanis2000}.  Over the lifetime of a
vortex, $t_{\rm life}$, the amount of dust the can be gathered by a
vortex is very large.  \cite{Chavanis2000} argues that this mass is
\begin{equation}
M_{\rm d} \sim \Omega t_{\rm life} \Sigma_{\rm d} R^2 f^2(t_{\rm stop}),
\end{equation}
where $f$ describes the efficiency of capturing dust and is $\sim 1$
when $t_{\rm stop} \sim t_{\rm dyn}$ and $R$ is the size scale of the
vortex.  If the inward concentration of dust is balanced by the
outward diffusion of this dust concentration due to turbulence, these
dust particles would be confined to a region on a scale of
\begin{equation} 
r_{\rm d} \sim \sqrt{Dt_{\rm capt}},
\end{equation}
where $D\sim \alpha R^2 \Omega$ is the turbulent diffusivity.  For
$\alpha = 0.01$, $r_{\rm d} \sim 0.1 R$, i.e., in the central core of
the vortex.  Hence the surface density is enhanced by two orders of
magnitude.  Since the GI hypothesis for planetesimal formation only
demands more modest increases, vortices should be ideal sites of
planetesimal formation.

However, such a increase in dust surface density is not without its
costs.  As we have shown a sufficient increase in the effective mean
molecular weight of the gas in the cores of vortices is destabilizing.
The condition for the HCI demands a mean molecular weight increase of
order 20\% or an increase of the dust surface density by a factor of
20\% if the dust and gas densities are similar in the midplane.  This
increase in dust surface density is much smaller than what is required
for the GI hypothesis, which demands a factor of a few increase
\citep{Youdin2002,ChiangYoudin2009} if the vertical structure of dust is
taken into account to a factor of $20-100$ when vertical structure is
ignored \citep{Chavanis2000}.  Thus, the HCI will be triggered before
gravitational instability sets in according to the present linear
calculation.  

There are many issues involving the stability of vortices with a heavy
core than cannot be resolved by the present linear calculation, which
we now briefly discuss. The first issue is the non-linear state of the
instability, which is not known at present.  We expect the HCI to grow
until its saturates, which may 1. destroy the vortex, 2. increase the
velocity dispersion of the dust layer, or 3. limit the enhancement in
the dust surface density to a few tens of percent, i.e., marginal
stability.  For any of these options, GI is curtailed in cores of
vortices.

Another issue is the equilibrium distribution of dust along a
streamline.  We have assumed that the dust is uniformly distributed
along a streamline.  For light particles, this is likely the case.
\cite{Tanga1996} and \cite{Chavanis2000} studied the process of dust
trapping in vortices and found that the zeroth order motion is that
light particles of dust travels along the elliptical streamlines with
a slow "radial" drift due to drag forces.  However, heavy particles
move along epicycles, i.e., ellipses with aspects ratio 2.  In
addition, \cite{Youdin2008} showed that the stationary point for dust
in a sub-Keplerian gas is not the center of the vortex, but rather a
point that is forward in azimuth.  This is unsurprising as the dust,
in maintaining a sub-Keplerian rotation rate, demands an additional
radial force away from the central star to counteract gravity, a force
that is supplied by gas pushing on the dust if the dust is ahead of
the vortex center in azimuth.  These elements suggest that the
distribution of dust along a streamline may not be uniform and so may
affect the stability properties of vortices in a non-trivial way.

A third issue is the nature of gas-dust coupling.  We have assumed the
gas and dust are well coupled on a dynamical time.  However, this may
not be the case.  For instance, the fastest settling dust is that
which is marginally coupled to the gas, i.e., $\Omega t_{\rm s} = 1$,
where $t_{\rm s}$ is the dust stopping time, \citep{Johansen2004}.
The dust that is trapped in vortices may be preferentially of a
certain size, i.e., marginally coupled to the gas.  Hence the
instability growth time, dynamical time, and dust-gas coupling time,
in the dusty protoplanetary disk can be all of the same order. 

Additional instabilities that arise directly from this gas-dust
coupling may also be important.  \cite{Youdin2005} showed that the
imperfect coupling between gas and dust and their backreaction on each
other leads to a secular streaming instability.  This instability
leads to protoplanetary disk turbulence and tends to concentrate dust
\citet{Youdin2007,Johansen2007}.  These streaming instabilities or
analogues may also be important in the stability protoplanetary
vortices and would be profitable to explore.  A proper accounting of
gas-dust coupling in a vortex and its effect on the VRTI is a topic of
future work.  

Another issue is that the vertical structure of the gas and the dust
is very different in protoplanetary disks.  Dust tends to settle
toward the midplane (though the presence of vortices and/or turbulence
may counter this tendency).  This settled dust may drive vertical
turbulence if it is sufficiently concentrated \citep[see for
  instance][]{Chiang2008,Barranco2009}.  The effect of this difference
in the vertical structure of gas and dust on vortices has not yet been
addressed and is likely important for both their equilibrium and
stability.  However, we may expect the HCI to be important regardless
because is a 2-D effect in a thin dust layer within a thicker gas
vortex.

Finally, the alert reader (and referee) will note that we have not discussed the case of $1 < \chi < 2$ vortices, i.e., low pressure vortices.  In principle, such vortices might exist in protoplanetary discs, but the prevailing theoretical bias is that vortices in protoplanetary discs are high pressure regions.  We have gone along with this bias in this work and have ignored these low pressure vortices.  However, we note that heavy core low pressure vortices are violently unstable (equivalent to the light core case discussed above) because of the reversal in the direction of the effective gravity.  In addition, it is unclear if these vortices would concentrate dust.  The equivalent calculation of \cite{Chavanis2000} for low pressure vortices has not been performed.  While a detailed study of low pressure protoplanetary vortices might be interesting, the impact of such a study is unclear.  

\section{Conclusions and Open Issues}\label{sec:conclusions}

We have demonstrated two instabilities in protoplanetary vortices,
resulting from light cores (VRTI) and sufficiently heavy cores (HCI).
The physics of the VRTI is analogous to the Rayleigh-Taylor
instability, with gravity replaced by centrifugal and centripetal
forces in a rotating fluid. The HCI appears to be a parametric
instability.  We have shown that these instabilities are robust for
all vortices possessing a light or sufficiently heavy core.  For
protoplanetary vortices, the instability of interest is the HCI as
dust would concentrate in their centers, leading to heavy cores.
While the nonlinear state of the these remains unexplored, we expect
that this instability prevents vortices from acting as protoplanetary
nurseries.

Both the VRTI and HCI are novel among elliptical vortex instabilities
as they are 2-D -- only motions in the x-y plane are required.
Previous work on the stability of vortices have focused on the
importance of the instabilities that involve 3-D motions
\citep{Lithwick2009,Lesur2009}.  Indeed, 3-D effects do lead to
additional instabilities that destroy vortices even before they can
collect dust. \cite{Lesur2009} argued that the 3-D elliptical
instability can destroy Kida vortices for $\chi < 4$ and $\chi > 6$.
\cite{Lithwick2009} argues that nonlinear coupling and transient
amplification between a vortex and its children that involve a
vertical component leads to destruction of the vortex.  He argues that
the stability requirement for any vortex is then that the base of the
vortex be larger (by at least a factor of 2) than its height.  Our
results suggest that if gas vortices having $\chi \gtrsim 6$ (as
proposed by \cite{Lithwick2009}), the HCI will affect these vortices
as they gather dust.

The analysis that we have attempted here is linear and so it is highly
dependent on the background equilibrium state.  The two equilibria
(Kida and GNG) that we have analyzed in this paper were chosen due to
their simple analytic structure.  Although we have shown that the HCI
is very similar in both these cases, 3-D simulations have clearly
demonstrated that vortices in protoplanetary disks are not so simple
\citet{Barranco2005,Shen2006,Lithwick2009}.

Finally, our work leaves open a number of issues including the
nonlinear state of the HCI, gas-dust coupling physics, and equilibrium
structure and vertical structure of vortices.  We are currently
pursuing numerical work exploring the effects of heavy vortices in
protoplanetary disk.

\acknowledgements

We thank Joe Barranco, Peter Goldreich, Denis Sipp, Yanquin Wu, and
Andrew Youdin for useful discussions. We thank the anonymous referee for useful comments. 
J.S.O. would like to thank Ed
Spiegel for inspiring this work. We would also like to thank the
staffs of Sugarlump and Ritual Roasters, where the majority of this
work was completed, for their hospitality and free wireless.  P.C. is
supported by the Canadian Institute for Theoretical Astrophysics.
J.S.O. is supported by NSF grant AST09-08553.

\appendix
\section{A Simple Example of Vortical Rayleigh-Taylor Instability}\label{sec:simple}

In this appendix, we discuss a simple example of the Vortical
Rayleigh-Taylor instability (VRTI) in a terrestrial vortex to
illustrate its basic physics.  Our simple treatment is derived from
\citet{Sipp2005}, and a detailed overview of the state of terrestrial
heavy vortex instability theory is found therein.
\begin{figure}
  \begin{center}
    \includegraphics[width=0.48\columnwidth]{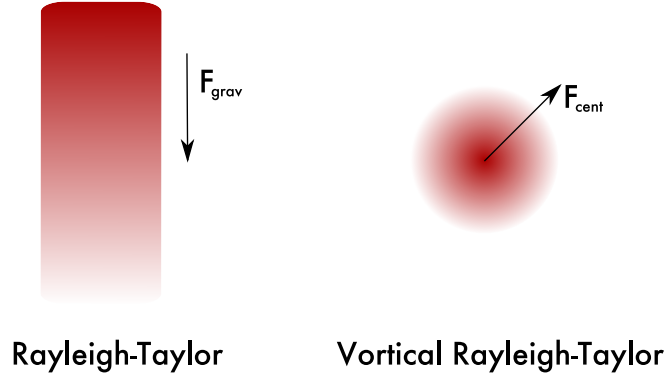}    
  \end{center}
  \caption{A cartoon showing the salient feature of the
    Rayleigh-Taylor type instability, which is that the density
    gradient (represented by the shading: dark tones correspond to
    high density, light tones to low density) is in the opposite
    direction from the restoring force. In the case of the standard
    Rayleigh-Taylor instability, the force is gravity; in the vortical
    Rayleigh-Taylor case it is the centrifugal force.}
  \label{f:VRTI_cartoon}
\end{figure}

For a 2-d circular vortex in equilibrium, pressure forces (inward) are
counterbalanced by centrifugal forces (outward).  Incompressible
motions of this 2-d vortex are described by the continuity equation,
\begin{equation}\label{eq:continuity_VRTI}
\ddt{\rho} + u\ddr{\rho} + \frac v r \ddphi{\rho} = 0,
\end{equation}
where $\rho$ is the density, $u = \dot{r}$ and $v = r\dot{\phi}$, the
momentum equations,
\begin{eqnarray}
\ddt{u} + u\ddr{u} + \frac v r \ddphi{u} - \frac {v^2} r &=& -\frac{1}{\rho}\ddr{P}, \\
\ddt{v} + u\ddr{v} + \frac v r \ddphi{v} + \frac {uv} r &=& -\frac{1}{\rho r}\ddphi{P},
\end{eqnarray}
where $P$ is the pressure, and the incompressibility condition,
\begin{equation}\label{eq:incompressible}
r^{-1}\ddr{ru} + \frac 1 r \ddphi{v} = 0.
\end{equation}

In equilibrium, the fluid motions of the vortex are circular and
constant, i.e., $u = 0$ and $v$ is constant.  Hence we find that
\begin{equation}
\rho^{-1}\ddr{P} = \frac {v^2} r.
\end{equation}
We now perturb equations
(\ref{eq:continuity_VRTI})-(\ref{eq:incompressible}) and assume
perturbations of the form $\exp(-i\omega t + im\phi + ikr)$.  The perturbed
continuity and momentum equations read
\begin{eqnarray}
-i\sigmabar\delta\rho + \delta u\ddr{\rho} &=& 0\label{eq:perturbed_rho}\\
-i\sigmabar\delta u - \frac {2v}{r}\delta v &=& -ik\frac{\delta P}{\rho} + 
\frac 1 {\rho} \ddr{P} \frac {\delta\rho}{\rho}\label{eq:perturbed_u}\\
-i\sigmabar\delta v + \left(\ddr{v} + \frac v r\right)\delta u &=& -i\frac{m}{r}\frac{\delta P}{\rho}\label{eq:perturbed_v},
\end{eqnarray}
where $\sigmabar = \left(\omega - {mv}/r \right)$.
The incompressibility condition (eq.[\ref{eq:incompressible}]) becomes 
\begin{equation}\label{eq:perturbed_incompressibility}
ik\delta u + \frac {im}{r} \delta v = 0,
\end{equation}
where we have assumed $kr \gg 1$. Equation
(\ref{eq:perturbed_incompressibility}) gives $\delta v$ in terms of
$\delta u$, which we apply to equations (\ref{eq:perturbed_u}) and
(\ref{eq:perturbed_v}).  Using equation (\ref{eq:perturbed_rho}) for
$\delta\rho$ in terms of $\delta u$, we find the dispersion relation:
\begin{equation}\label{eq:dispersion_VRTI}
\sigmabar^2\left(\frac {m^2}{r^2} + k^2\right) + i\sigmabar k\frac {m}{r}\left(\ddr{v} - \frac v r\right) - \frac {m^2}{r^2} \frac {v^2}{r}\ddr{\ln\rho} = 0.
\end{equation}

For an uniformly rotating vortex, $v\propto r$, the second term in
(\ref{eq:dispersion_VRTI}) vanishes and the solution to the dispersion
relation is
\begin{equation}\label{eq:simple_dispersion}
\sigmabar^2 = \frac {m^2/r^2}{m^2/r^2 + k^2} \frac {v^2}{r}\ddr{\ln\rho},
\end{equation}
which is $<0$ (unstable) if $\partial{\ln\rho}/\partial r < 0$, that
is, if the core of the vortex is heavy.\footnote{This does not violate
  the Rayleigh criterion, which states that flows with ${d v^2
    r^2}/{dr} > 0$ are stable to axisymmetric perturbations, as these
  perturbations are non-axisymmetric.}  This instability is analogous
to the Rayleigh-Taylor instability, whose dispersion relation is
$\omega \propto k_{\perp}^2/k^2$, but where gravity is replaced by a
centrifugal force.  In the case of the Rayleigh-Taylor instability, the
equilibrium is set by pressure forces balancing gravity.  Heavy fluid
that sits on top of light fluid which fulfills the conditions of
equilibrium, but is unstable to interpenetration across the interface.
By analogy, in the VRTI case, the equilibrium vortex is set by
pressure forces balancing centrifugal forces.  The presence of a heavy
core leads to non-axisymmetric instabilities (where the origin is set
by the center of the vortex), where again the heavy fluid elements in
the core interpenetrate light fluid on the
exterior. Figure~\ref{f:VRTI_cartoon} illustrates this analogy.

The analogy is made clearer if we identify $k_{\phi} \equiv m/r$,
which is perpendicular to the centrifugal force (i.e., the effective
gravity), which is in the $\hat{r}$ direction.  Thus, we can make the
identification $k_{\phi}^2/({k_{\phi}^2 + k^2}) \leftrightarrow
k_{\perp}^2/k^2$ between VRTI and RTI, respectively.  The wavevector
that grows the fastest in both instabilities is the wavevector that is
perpendicular to the vertical gravity and the radially outward
centrifugal force, respectively.

\bibliographystyle{apj} 
\bibliography{ms}

\end{document}